\documentstyle[nato]{crckapb}

\begin{opening}

\title{SUPERSYMMETRIC HYBRID INFLATION}

\author{G. LAZARIDES}

\institute{Physics Division, School of Technology, \\ 
Aristotle University of Thessaloniki, \\
Thessaloniki GR 540 06, Greece}

\end{opening}

\runningtitle{SUPERSYMMETRIC HYBRID INFLATION}

\begin{document}

\begin{abstract}
The non-supersymmetric and supersymmetric versions of 
hybrid inflation are summarized. In the latter, the 
necessary inclination along the inflationary trajectory 
is provided by radiative corrections. Supersymmetric hybrid 
inflation (with its extensions) is an extremely `natural' 
inflationary scenario. The reasons are that it does not 
require `tiny' parameters, its superpotential has the most 
general form allowed by the symmetries, and it can be 
protected  against radiative or supergravity corrections. 
Concrete supersymmetric grand unified theories which lead 
to hybrid inflation, solve the $\mu$ problem via a 
Peccei-Quinn symmetry and generate seesaw masses for the 
light neutrinos can be constructed. As an example, we 
present a theory with unified gauge group $SU(3)_c\times 
SU(2)_L\times SU(2)_R\times U(1)_{B-L}$. The `reheating' 
which follows hybrid inflation is studied. It is shown that 
the gravitino constraint on the `reheat' temperature can be 
`naturally' satisfied. Also, the observed baryon asymmetry 
of the universe can be generated via a primordial 
leptogenesis consistently with the requirements from solar
and atmospheric neutrino oscillations. Extensions of the 
standard supersymmetric hybrid inflationary scenario which 
are still consistent with all these requirements but can 
also avoid the cosmological disaster from the possible 
copious monopole production at the abrupt termination of 
standard hybrid inflation are constructed. They rely on 
utilizing the leading non-renormalizable correction to the 
standard hybrid inflationary superpotential and are 
necessary for higher unified gauge groups such as $SU(4)_c
\times SU(2)_L\times SU(2)_R$ which predict the existence 
of monopoles. In one extension, which we call shifted hybrid 
inflation, the relevant part of inflation takes place along 
a `shifted' classically flat direction on which the unified 
gauge symmetry is already broken. In the other extension, 
called smooth hybrid inflation, the trilinear term of the 
standard hybrid inflationary superpotential is removed by a 
discrete symmetry. The inflationary path then possesses a 
classical inclination and the termination of inflation is 
smooth.

\end{abstract}

\section{Hybrid Inflation} 
\label{sec:hybrid}

\subsection{The non-Supersymmetric Version} 
\label{subsec:nonsusy}

The most important disadvantage of inflationary scenarios 
such as the `new' \cite{new} or `chaotic' \cite{chaotic} 
ones is that they require `tiny' coupling constants in order 
to reproduce the measurements of the cosmic background 
explorer (COBE) \cite{cobe} on the cosmic microwave 
background radiation (CMBR). This difficulty was overcome 
by Linde \cite{hybrid} who proposed, in the context of 
non-supersymmetric grand unified theories (GUTs), the hybrid 
inflationary scenario. The basic idea was to use two real 
scalar fields $\chi$ and $\sigma$ instead of one that was 
normally used. $\chi$ provides the `vacuum' energy density 
which drives inflation, while $\sigma$ is the slowly varying 
field during inflation. This splitting of roles between two 
fields allows us to reproduce the observed temperature 
fluctuations of the CMBR with `natural' (not too small) 
values of the relevant parameters in contrast to previous 
realizations of inflation. 

\par
The scalar potential utilized by Linde is
\begin{equation}
V(\chi,\sigma)=\kappa^2 \left(M^2-
\frac{\chi^2}{4}\right)^2+
\frac{\lambda^2\chi^2 \sigma^2}{4}
+\frac {m^2\sigma^2}{2}~,
\label{eq:lindepot}
\end{equation}
where $\kappa$, $\lambda$ are dimensionless positive 
coupling constants and $M$, $m$ are mass parameters. The 
vacua lie at $\langle\chi\rangle=\pm 2M$, 
$\langle\sigma\rangle=0$. Putting $m$=0, for the moment, 
we observe that $V$ possesses an exactly flat direction at 
$\chi=0$ with $V(\chi=0,\sigma)=\kappa^2M^4$. The 
${\rm mass}^2$ of the field $\chi$ along this flat direction 
is $m^2_\chi=-\kappa^2 M^2+\lambda^2\sigma^2/2$. So, for  
$\chi=0$ and $\vert\sigma\vert>\sigma_c=\sqrt{2}\kappa M/
\lambda$, we obtain a flat valley of minima. Reintroducing 
$m\neq 0$, this valley acquires a non-zero slope and the 
system can inflate as it rolls down this valley. This 
scenario is called hybrid since the `vacuum' energy density 
($=\kappa^2M^4$) is provided by $\chi$, while the slowly 
rolling field (inflaton) is $\sigma$. 

\par
The $\epsilon$ and $\eta$ criteria (see e.g., 
Ref.\cite{cosmology}) imply that, for the relevant values 
of parameters (see below), inflation continues until 
$\sigma$ reaches $\sigma_c$, where it terminates abruptly. 
It is followed by a `waterfall', i.e., a sudden entrance into 
an oscillatory phase about a global minimum. Since the system 
can fall into either of the two minima with equal probability, 
topological defects (monopoles, walls or cosmic strings) are 
copiously produced \cite{smooth} if they are predicted by the 
particular particle physics model employed. So, if the 
underlying GUT gauge symmetry breaking (by 
$\langle\chi\rangle$) leads to the existence of monopoles 
or walls, we encounter a cosmological catastrophe.

\par
The onset of hybrid inflation requires \cite{onset} that, 
at a cosmic time of order $H^{-1}$, $H$ being the Hubble 
parameter during inflation, a region exists in the universe 
with size greater than about $H^{-1}$, where $\chi$ and 
$\sigma$ happen to be almost uniform with negligible kinetic 
energies and values close to the bottom of the valley of 
minima. Such a region, at the Planck time $t_P=M_P^{-1}$ 
($M_P\approx 1.22\times 10^{19}~{\rm GeV}$ is the Planck 
mass), would have been much larger than the Planck length 
$\ell_P=M_P^{-1}$ and it is, thus, very difficult to imagine 
how it could emerge so homogeneous. Moreover, as it has been 
argued \cite{initial}, the initial values (at $t_P$) of the 
fields in this region must be strongly restricted in 
order to obtain adequate inflation. Several possible 
solutions to this problem of initial conditions for hybrid 
inflation have been already proposed (see e.g., 
Refs.\cite{double,sugra,costas}). 

\par
The quadrupole anisotropy of CMBR produced during hybrid 
inflation can be estimated, using the standard formulae 
(see e.g., Ref.\cite{cosmology}), to be 
\begin{equation}
\left(\frac{\delta T}{T}\right)_{Q}\approx 
\left(\frac{16\pi}{45}\right)^{\frac{1}{2}} 
\frac{\lambda\kappa^2M^5}{M^3_Pm^2}~\cdot
\label{eq:lindetemp}
\end{equation}
The COBE \cite{cobe} result, 
$(\delta T/T)_{Q}\approx 6.6\times 10^{-6}$, can then be 
reproduced with $M\approx 2.86\times 10^{16}~{\rm GeV}$, 
the supersymmetric (SUSY) GUT vacuum expectation value (vev), 
and $m\approx 1.3~\kappa\sqrt{\lambda}\times 
10^{15}~{\rm GeV}$. Note that $m\sim 10^{12}~{\rm GeV}$ 
for $\kappa$, $\lambda\sim 10^{-2}$. 

\subsection{The Supersymmetric Version}
\label{subsec:susy}

Hybrid inflation turns out \cite{lyth} to be `tailor made' for 
application to globally SUSY GUTs except that an intermediate 
scale mass for $\sigma$ cannot be obtained in this context. 
Actually, all scalar fields acquire masses of order 
$m_{3/2} \sim 1~{\rm TeV}$ (the gravitino mass) from soft 
SUSY breaking.

\par 
Let us consider the renormalizable superpotential
\begin{equation} 
W=\kappa S(-M^2+\bar{\phi}\phi),
\label{eq:superpot}
\end {equation}
where $\bar{\phi}$, $\phi$ is a conjugate pair of $G_{S}$ 
(the standard model gauge group) singlet left handed 
superfields belonging to non-trivial representations of the 
GUT gauge group $G$ and reducing its rank by their vevs, and 
$S$ is a gauge singlet left handed superfield. The 
parameters $\kappa$ and $M$ ($\sim 10^{16}~{\rm GeV}$) 
can be made positive by field redefinitions. The vanishing 
of the F-term $F_S$ implies that 
$\langle\bar{\phi}\rangle\langle\phi\rangle=M^2$, 
whereas the D-terms vanish for $\vert\langle\bar{\phi}
\rangle\vert=\vert\langle\phi\rangle\vert$. So, the 
SUSY vacua lie at $\langle\bar{\phi}\rangle^*=
\langle\phi\rangle=\pm M$ and $\langle S\rangle=0$ 
(from $F_{\bar{\phi}}=F_{\phi}=0$). We see that $W$ 
leads to the spontaneous breaking of $G$.

\par
The same superpotential $W$ gives rise to hybrid inflation. 
The potential derived from $W$ in Eq.(\ref{eq:superpot}) is
\begin{equation}
V(\bar{\phi},\phi,S)=
\kappa^2\vert M^2-\bar{\phi}\phi\vert^2+
\kappa^2\vert S\vert^2(\vert\bar{\phi}\vert^2+
\vert\phi\vert^2)+{\rm{D-terms}}.
\label{eq:hybpot}
\end{equation}
D-flatness implies $\bar{\phi}^*=e^{i\theta}\phi$. We 
take $\theta=0$, so that the SUSY vacua are contained. Note 
that $W$ possesses a $U(1)_R$ R-symmetry: 
$\bar{\phi}\phi\to\bar{\phi}\phi$, $S\to e^{i\alpha}S$, 
$W\to e^{i\alpha}W$. Actually, $W$ is the most general 
renormalizable superpotential allowed by $U(1)_R$ and $G$. 
Performing appropriate $G$ and R-transformations, we bring 
$\bar{\phi}$, $\phi$, $S$ on the real axis, i.e., 
 $\bar{\phi}=\phi\equiv\chi/2$, $S\equiv\sigma/\sqrt{2}$
where $\chi$, $\sigma$ are normalized real scalar fields. 
$V$ then takes the form in Eq.(\ref{eq:lindepot}) with 
$\kappa=\lambda$ and $m=0$. So, Linde's potential for hybrid 
inflation is almost obtainable from SUSY GUTs but without the 
mass term of $\sigma$ which is, however, crucial for driving 
the inflaton towards the vacua.

\par
One way to generate the necessary slope along the inflationary 
trajectory is \cite{dss} to include the one-loop radiative 
corrections on this trajectory ($\bar{\phi}=\phi=0$, 
$\vert S\vert>S_{c}\equiv M$). In fact, SUSY breaking by 
the `vacuum' energy density $\kappa^2M^4$ along this valley 
causes a mass splitting in the supermultiplets $\bar{\phi}$, 
$\phi$. We obtain a Dirac fermion with ${\rm mass}^2$ equal 
to $\kappa^2\vert S\vert^2$ and two complex scalars with 
${\rm mass}^2$ equal to 
$\kappa^2\vert S\vert^2\pm\kappa^2M^2$. This leads to the 
existence of important one-loop radiative corrections to $V$ 
on the inflationary valley which can be found from the 
Coleman-Weinberg formula \cite{cw}:
\begin{equation}
\Delta V=\frac{1}{64\pi^2}\sum_i(-)^{F_i}\ M_i^4\ln
\frac{M_i^2}{\Lambda^2}~, 
\label{eq:deltav}
\end{equation}
where the sum extends over all helicity states $i$, $F_i$ and
$M_i^2$ are the fermion number and ${\rm mass}^2$ of the 
$i$th state, and $\Lambda$ is a renormalization mass scale. 
We find that $\Delta V(\vert S\vert)$ is given by
\begin{equation}
\kappa^2 M^4~{\kappa^2N\over 32\pi^2}\left(
2\ln{\kappa^2\vert S\vert^2\over\Lambda^2}
+(z+1)^{2}\ln(1+z^{-1})+(z-1)^{2}\ln(1-z^{-1})\right),
\label{eq:rc}
\end{equation}
where $z=x^2=\vert S\vert^2/M^2$ and $N$ is the 
dimensionality of the representations to which $\bar{\phi}$, 
$\phi$ belong. For $z\gg 1$ ($\vert S\vert\gg S_c$), the 
effective potential on this trajectory can be expanded as 
\cite{dss,lss}
\begin{equation}
V_{{\rm{eff}}}(\vert S\vert)=\kappa^2 M^4
\left[1+\frac{\kappa^2N}{16\pi^2}\left(\ln 
\frac{\kappa^2\vert S\vert^2}{\Lambda^2}
+\frac{3}{2}-\frac{1}{12z^2}+\cdots\right)\right].
\label{eq:veff}
\end{equation}
Note that the slope along the inflationary valley which is 
provided by these radiative corrections is 
$\Lambda$-independent.

\par
From Eq.(\ref{eq:rc}) and using the standard formalism 
(see e.g., Ref.\cite{cosmology}), we find that the 
quadrupole anisotropy of CMBR is
\begin{equation} 
\left(\frac{\delta T}{T}\right)_{Q}\approx\frac{8\pi}
{\sqrt{N}}\left(\frac{N_{Q}}{45}\right)^{\frac{1}{2}} 
\left(\frac{M}{M_{P}}\right)^2x_Q^{-1}y_Q^{-1}
\Lambda(x_Q^2)^{-1},
\label{eq:qa} 
\end{equation}
with
\begin{equation}
\Lambda(z)=(z+1)\ln(1+z^{-1})+(z-1)\ln(1-z^{-1})~, 
\label{eq:lambda} 
\end{equation}
\begin{equation}
y_Q^2=\int_1^{x_Q^2}\frac{dz}{z}\Lambda(z)^{-1}, 
~y_Q\geq 0.
\label{eq:yq} 
\end{equation}
Here, $N_{Q}$ is the number of e-foldings suffered by our 
present horizon scale during inflation, and 
$x_Q=\vert S_Q\vert/M$, with $S_Q$ being the value of 
$S$ when our present horizon scale crossed outside the 
inflationary horizon. For $\vert S_Q\vert\gg S_c$, $y_Q=
x_Q(1-7/12x_Q^2+\cdots)$. Finally, from 
Eq.(\ref{eq:rc}), one finds
\begin{equation} 
\kappa\approx\frac{8\pi^{\frac{3}{2}}}{\sqrt{NN_Q}}
~y_Q~\frac{M}{M_{P}}~\cdot
\label{eq:kappa}
\end{equation}  

\par
The slow roll conditions (see e.g., Ref.\cite{cosmology}) for 
SUSY hybrid inflation are $\epsilon,\vert\eta\vert\ll 1$, 
where
\begin{equation}
\epsilon=\left(\frac{\kappa^2M_P}{16\pi^2M}\right)^2
\frac{N^2x^2}{8\pi}\Lambda(x^2)^2,
\label{eq:epsilon}
\end{equation}
\begin{equation}
\eta=\left(\frac{\kappa M_P}{4\pi M}\right)^2
\frac{N}{8\pi}\left((3z+1)\ln(1+z^{-1})+
(3z-1)\ln(1-z^{-1})\right).   
\label{eq:eta}
\end{equation} 
Note that $\eta\rightarrow-\infty$ as $x\rightarrow 1^+$.
However, for most relevant values of the parameters 
($\kappa\ll 1)$, the slow roll conditions are violated only 
`infinitesimally' close to the critical point at $x=1$ 
($\vert S\vert=S_c$). So, inflation continues practically 
until this point is reaches, where the `waterfall' occurs.

\par
From the COBE \cite{cobe} result, 
$(\delta T/T)_{Q}\approx 6.6\times 10^{-6}$, and 
eliminating $x_Q$ between Eqs.(\ref{eq:qa}) and 
(\ref{eq:kappa}), we obtain $M$ as a function of $\kappa$. 
For $x_Q\rightarrow\infty$, $y_Q\rightarrow x_Q$ 
and $x_Qy_Q\Lambda(x_Q^2)\rightarrow 1^-$. Thus, the 
maximal $M$ is achieved in this limit and equals about 
$10^{16}~{\rm GeV}$ (for $N=8$, $N_Q\approx 55)$. This value 
of $M$, although somewhat smaller than the SUSY GUT scale, is 
quite close to it. As a numerical example, take $\kappa=4\times 
10^{-3}$ which gives $M\approx 9.57\times 10^{15}~{\rm GeV}$, 
$x_Q\approx2.633$, $y_Q\approx 2.42$. The slow roll conditions 
are violated at $x-1\approx 7.23\times 10^{-5}$, where 
$\eta=-1$ ($\epsilon\approx 8.17\times 10^{-8}$ at $x=1$). 
The spectral index of density perturbations $n=1-6\epsilon+2\eta$
\cite{liddle} is $\approx 0.985$.  

\par
The SUSY hybrid inflationary scenario can be considered `natural' 
for the following reasons:

\begin{list}
\setlength{\rightmargin=0cm}{\leftmargin=0cm}

\item[{\bf i.}] 
There is no need of `tiny' coupling constants 
($\kappa\sim 10^{-3}-10^{-2}$).
\vspace{.25cm}
\item[{\bf ii.}] 
The superpotential $W$ in Eq.(\ref{eq:superpot}) has the 
most general renormalizable form allowed by the gauge and 
R-symmetries. Moreover, the coexistence of the $S$ and 
$S\bar{\phi}\phi$ terms in $W$ implies that the combination 
$\bar{\phi}\phi$ is `neutral' under all possible symmetries
of $W$ and, thus, all the non-renormalizable terms of the form
$S(\bar{\phi}\phi)^n$, $n\geq 2$, are also necessarily 
present in $W$ \cite{jean}. The leading term of this type 
$S(\bar{\phi}\phi)^2$, if its dimensionless coefficient is 
of order unity, can be comparable to $S\bar{\phi}\phi$ 
(recall that $\kappa\sim 10^{-3}$) and, thus, play an 
important role in inflation (see Sec.\ref{sec:extensions}). 
All higher order terms of this type with $n\geq 3$ give 
negligible contributions to the inflationary potential. The 
presence of $U(1)_R$ is crucial since it guarantees the 
linearity of $W$ in $S$ to all orders excluding terms such 
as $S^2$ which could generate an inflaton mass $\geq H$, 
thereby ruining inflation by violating the slow roll 
conditions.
\vspace{.25cm}
\item[{\bf iii.}] 
SUSY guarantees that the radiative corrections do not 
invalidate \cite{susyinfl} inflation, but rather provide 
\cite{dss} a slope along the inflationary trajectory, 
needed for driving the inflaton towards the SUSY vacua.
\vspace{.25cm}
\item[{\bf iv.}] 
Supergravity (SUGRA) corrections can be brought under control 
leaving inflation intact. The scalar potential in SUGRA is 
given \cite{superg} by
\begin{equation}
V=\exp\left(\frac{K}{m_P^2}\right)\left[\left(
K^{-1}\right)_i^{~j}F^i F_j-3\frac{\left|W\right|^2}
{m_P^2}\right],
\label{eq:sugra}
\end{equation} 
where $K$ is the K\"ahler potential, $m_P=M_P/\sqrt{8\pi}
\approx 2.44\times 10^{18}~{\rm GeV}$ in the `reduced' 
Planck scale, $F^i=W^i+K^iW/m_P^2$, and upper (lower) 
indices denote differentiation with respect to the scalar 
field $\phi_i$ ($\phi^{j*}$). $K$ can be expanded as $K=
\vert S\vert^2+\vert\bar{\phi}\vert^2+\vert\phi\vert^2+
\alpha\vert S\vert^4/m_P^2+\cdots$, where the leading 
(quadratic) terms constitute the `minimal' K\"ahler 
potential. The term $\vert S\vert^2$, whose coefficient is 
necessarily normalized to unity, could generate a 
${\rm mass}^2\sim\kappa^2M^4/m_P^2\sim H^2$ for $S$ 
along the inflationary trajectory from the expansion of the 
exponential prefactor in Eq.(\ref{eq:sugra}). This would 
ruin inflation. Fortunately, with this particular form of 
$W$ (including all the higher order terms) this 
${\rm mass}^2$ is exactly cancelled in $V$ 
\cite{lyth,stewart}. The linearity of $W$ in $S$, which 
is guaranteed to all orders by the R-symmetry, is crucial 
for this cancellation to take place. This is an important 
property of this scheme. The $\vert S\vert^4$ term in 
$K$ also generates a ${\rm mass}^2$ for $S$ via the factor 
$(\partial^2 K/\partial S\partial S^*)^{-1}=1-4\alpha
\vert S\vert^2/m_P^2+\cdots$ in Eq.(\ref{eq:sugra}), 
which is however not cancelled (see e.g., Ref.\cite{quasi}). 
In order to avoid ruining the present inflationary scheme, 
one has then to assume \cite{sugra,lss} 
that $|\alpha|\stackrel{_{<}}{_{\sim }}10^{-3}$. All 
other higher order terms in $K$ are harmless since they 
give suppressed contributions ($\vert S\vert\ll m_P$ on 
the inflationary path). So, we see that a mild tuning of 
just one parameter is adequate for controlling SUGRA 
corrections. This is a great advantage of this model since 
in other cases tuning of infinitely many parameters is 
required. Moreover, note that with special forms 
of the K\"{a}hler potential one can solve this problem 
even without a mild tuning. An example is given in 
Ref.\cite{costas}, where the dangerous ${\rm{mass}}^2$ 
could be cancelled in the presence of fields without 
superpotential but with large vevs generated via D-terms. 
This property practically persists even in the extensions 
of the model we will consider in Sec.\ref{sec:extensions}.
\end{list} 
In summary, for all these reasons, we consider SUSY hybrid 
inflation (with its extensions) as an extremely `natural' 
inflationary scenario.

\section{Hybrid Inflation in Concrete SUSY GUTs} 
\label{sec:guts}

We will now discuss how the SUSY hybrid inflationary scenario 
can be `embedded' in concrete SUSY GUTs. As an example, let us 
consider a moderate extension of the minimal supersymmetric 
standard model (MSSM) based on the left-right symmetric gauge 
group
\begin{equation}
G_{LR}=SU(3)_c\times SU(2)_L\times SU(2)_R
\times U(1)_{B-L}
\label{eq:lr}
\end{equation} 
(see Refs.\cite{lss,hier,trieste}). The breaking of 
$G_{LR}$ to $G_S$ is achieved via a conjugate pair of 
$SU(2)_R$ doublet superfields $\bar l^{c}$, $l^c$ with 
$B-L$ (baryon minus lepton number) equal to -1, 1, which 
acquire vevs along their right handed neutrino directions 
$\bar\nu^{c}_{H}$, $\nu^c_H$ corresponding to the 
superfields $\bar{\phi}$, $\phi$ in Sec.\ref{subsec:susy}. 
The (renormalizable) superpotential for the breaking of 
$G_{LR}$ is
\begin{equation}
W=\kappa S(-M^2+\bar l^{c}l^c), 
\label{eq:W}
\end{equation}
where $\kappa$, $M$ can be made positive by field redefinitions. 
This superpotential leads to hybrid inflation exactly as $W$ in 
Eq.(\ref{eq:superpot}). The quadrupole anisotropy of CMBR, 
$(\delta T/T)_Q$, and the coupling constant $\kappa$ are 
given by Eqs.(\ref{eq:qa}) and (\ref{eq:kappa}) with $N=2$ 
since $\bar l^{c}$, $l^c$ have two components each.

\par
An important shortcoming of MSSM is that there is no 
understanding of how the SUSY $\mu$ term, with the right 
magnitude of $|\mu|\sim 10^{2}-10^{3}~{\rm GeV}$, arises. 
One way \cite{rsym} to solve this $\mu$ problem is via a 
Peccei-Quinn (PQ) symmetry $U(1)_{PQ}$ \cite{pq}, which also 
solves the strong CP problem. This solution is based on the 
observation \cite{kn} that the axion decay constant $f_{a}$, 
which is the symmetry breaking scale of $U(1)_{PQ}$, is 
(normally) `intermediate' ($\sim 10^{11}-10^{12}~{\rm GeV}$) 
and, thus, $|\mu|\sim f_{a}^2/m_P$. The scale $f_{a}$ is, 
in turn, $\sim (m_{3/2}m_P)^{1/2}$, where 
$m_{3/2}\sim 1~{\rm{TeV}}$ is 
the gravity-mediated soft SUSY breaking scale (gravitino mass). 
In order to implement this solution of the $\mu$ problem, we 
introduce a pair of gauge singlet superfields $\bar{N}$, $N$ 
with PQ charges 1, -1 and the non-renormalizable couplings 
$\lambda_1N^2h^2/m_P$, $\lambda_2\bar{N}^2N^2/m_P$ in 
the superpotential. Here, $h=(h^{(1)}, h^{(2)})$ is the 
electroweak Higgs superfield, which is a bidoublet under 
$SU(2)_L\times SU(2)_R$, and $\lambda_{1,2}$ are taken 
positive by redefining the phases of $\bar{N}$, $N$. After 
SUSY breaking, the $\bar N^2N^2$ term leads to the scalar 
potential:
\begin{eqnarray}
V_{PQ}&=&\left(m_{3/2}^2
+4\lambda_2^2\left|\frac{\bar{N}N}{m_P}\right|^2\right)
\left[(|\bar{N}|-|N|)^2+2|\bar{N}||N|\right]
\nonumber \\
& &+2|A|m_{3/2}\lambda_2\frac{|\bar{N}N|^2}{m_P}
{\rm{cos}}(\epsilon+2\bar{\theta}+2\theta),
\label{eq:pqpot}
\end{eqnarray} 
where $A$ is the dimensionless coefficient of the soft SUSY 
breaking term corresponding to the superpotential term 
$\bar{N}^2N^2$ and $\epsilon$, $\bar{\theta}$, $\theta$  
are the phases of $A$, $\bar{N}$, $N$ respectively. 
Minimization of $V_{PQ}$ then requires 
$|\bar{N}|=|N|$, $\epsilon+2\bar{\theta}+2\theta=\pi$ 
and $V_{PQ}$ takes the form
\begin{equation}
V_{PQ}=2|N|^2m_{3/2}^2\left(4\lambda_2^2\frac{|N|^4}
{m_{3/2}^2m_P^2}-|A|\lambda_2\frac{|N|^2}{m_{3/2}m_P}
+1\right).
\label{eq:pqpotmin}
\end{equation}
For $|A|>4$, the absolute minimum of the potential is at
\begin{equation}
|\langle\bar{N}\rangle|=|\langle N\rangle|\equiv
\frac{f_a}{2}=(m_{3/2}m_P)^{\frac{1}{2}}
\left(\frac{|A|+(|A|^2-12)^{\frac{1}{2}}}
{12\lambda_2}\right)^{\frac{1}{2}}\sim (m_{3/2}m_{P})
^{\frac{1}{2}}.
\label{eq:solution}
\end{equation}
The $\mu$ term is generated via the $N^2h^2$ superpotential 
term with $|\mu|=2\lambda_1|\langle N\rangle|^2/m_P$, 
which is of the right magnitude.

\par
The potential $V_{PQ}$ also has a local minimum at 
$\bar{N}=N=0$, which is separated from the global PQ minimum by 
a sizable potential barrier preventing a successful transition 
from the trivial to the PQ vacuum. This situation persists at all 
cosmic temperatures after the `reheating' which follows hybrid 
inflation, as has been shown \cite{jean} by considering the 
one-loop temperature corrections \cite{jackiw} to the potential. 
We are, thus, obliged to assume that, after the end of inflation, 
the system emerges in the PQ vacuum since, otherwise, it will be 
stuck for ever in the trivial vacuum.

\par
The gauge group $G_{LR}$ implies the presence of right handed 
neutrino superfields $\nu^c_i$ ($i=1,2,3$ is the family index), 
which form $SU(2)_R$ doublets $L^c_i=(\nu^c_i,e^c_i)$ with 
the $SU(2)_L$ singlet charged antileptons $e^c_i$. In order to 
generate `intermediate' scale masses for the $\nu^c$'s, we 
introduce the non-renormalizable superpotential couplings 
$\gamma_i\bar l^{c}\bar l^{c}L^c_iL^c_i/m_P$ (in a basis 
with diagonal and positive $\gamma$'s). The $\nu^c$ masses 
are then $M_i=2\gamma_iM^2/m_P$ (with 
$\langle\bar l^{c}\rangle$, $\langle l^{c}\rangle$ taken 
positive by a $B-L$ transformation). Light neutrinos acquire 
hierarchical masses via the seesaw mechanism and, thus, cannot 
play the role of hot dark matter (HDM) in the universe. They 
are more appropriate for a universe with non-zero cosmological 
constant ($\Lambda\neq 0$) favored by recent observations 
\cite{lambda}. The presence of HDM in such a universe is not 
necessary \cite{lambdafit,primack}. Note that the couplings 
which generate the $\nu^c$ masses are also responsible for 
the inflaton decay (see Sec.\ref{subsec:reheat}).

\par
From our discussion above, it is clear that the superpotential 
of the model must contain the following extra couplings:
\begin{equation}
N^{2}h^{2},~\bar{N}^2N^{2},~hQQ^c,~hLL^c,
~\bar l^{c}\bar l^{c}L^{c}L^{c}.
\label{eq:couplings}
\end{equation}
Here $Q_i$ and $L_i$ are the $SU(2)_L$ doublet left handed 
quark and lepton superfields, whereas $Q_i^c=(u^c_i,d^c_i)$ 
are the $SU(2)_R$ doublet antiquarks. The quartic terms in 
Eq.(\ref{eq:couplings}) carry a factor $m^{-1}_P$ which has 
been left out together with the dimensionless coupling constants 
and family indices.

\par
The continuous global symmetries of this superpotential are 
$U(1)_B$ (and, consequently, $U(1)_L$) with the extra 
superfields $\bar l^{c}$, $l^{c}$, $S$, $\bar{N}$, $N$ 
carrying zero baryon number, an anomalous PQ symmetry 
$U(1)_{PQ}$, and a non-anomalous R-symmetry $U(1)_R$. The 
$PQ$ and $R$ charges of the superfields are as follows 
($W$ carries one unit of $R$ charge):
\begin{eqnarray*}
PQ:~\bar l^{c},~l^{c},~S,~Q^c,~L^c~(0),~h,~\bar{N}~(1),
~Q,~L,~N~(-1);
\end{eqnarray*}
\begin{equation}
R:~\bar l^{c},~l^{c},~h,~\bar{N}~(0),~S~(1),~Q,~Q^c,~L,
~L^c,~N~(1/2).
\label{eq:charges}
\end{equation}
Note that $U(1)_B$ (and, thus, $U(1)_L$) invariance is 
automatically implied by $U(1)_R$ even if all possible 
non-renormalizable terms are included in the superpotential. 
This is due to the fact that the $R$ charges of the products 
of any three color (anti)triplets exceed unity and cannot be 
compensated since there are no negative $R$ charges 
available. 

\par
In order to avoid undesirable mixing of the components of 
the $L$'s with the ones of $\bar l^c$ or $h^{(2)}$ via the 
allowed superpotential couplings $\bar{N}NL^c\bar l^c$, 
$\bar{N}NLhl^c$, we impose a $Z_2$ symmetry 
(`lepton parity') under which $L$, $L^c$ change sign. This 
symmetry is equivalent to $Z_2$ `matter parity' under which 
$L$, $L^c$, $Q$, $Q^c$ change sign 
since $U(1)_B$ is also present and contains `baryon parity' 
under which $Q$, $Q^c$ change sign. The only superpotential 
terms which are permitted by $U(1)_{PQ}$, $U(1)_R$ and 
`matter parity' are the ones already included and 
$LLl^cl^c\bar{N}^2\bar l^cl^c$ and $LLl^cl^chh$ modulo 
arbitrary multiplications by non-negative powers of the 
combination $\bar l^cl^c$. The vevs of $\bar l^c$, $l^c$,
$\bar{N}$, $N$ leave unbroken only the symmetries $G_S$, 
$U(1)_B$ and `matter parity'. 

\subsection{`Reheating' and the Gravitino Constraint}
\label{subsec:reheat}

A complete inflationary scenario should be followed by a 
successful `reheating' satisfying the gravitino constraint 
\cite{khlopov} on the `reheat' temperature, $T_r
\stackrel{_{<}}{_{\sim}}10^9~{\rm GeV}$, and 
generating the observed baryon asymmetry of the universe 
(BAU). After the end of inflation, the system falls towards 
the SUSY vacuum and performs damped oscillations about it. 
The inflaton (oscillating system) consists of the two 
complex scalar fields $\theta=(\delta\bar\nu^c_H+
\delta\nu^c_H)/\sqrt{2}$ ($\delta\bar\nu^c_H=
\bar\nu^c_H-M$, $\delta\nu^c_H=\nu^c_H-M$) and $S$, 
with equal mass $m_{\rm infl}=\sqrt{2}\kappa M$. 

\par
The oscillating fields $\theta$ and $S$ decay into a pair of 
right handed neutrinos ($\psi_{\nu^c_i}$) and sneutrinos 
($\nu^c_i$) respectively via the superpotential 
couplings $\bar{l}^c\bar{l}^c L^cL^c$ and $S\bar{l}^cl^c$. 
The relevant Lagrangian terms are:
\begin{equation}
L^\theta_{\rm decay}=-\sqrt{2}\gamma_i\frac{M}{m_P}
\theta\psi_{\nu_i^c}\psi_{\nu_i^c}+ h.c.~, 
\label{eq:thetadecay}
\end{equation}
\begin{equation}
L^S_{\rm decay}=-\sqrt{2}\gamma_i\frac{M}{m_P}S^*\nu^c_i
\nu^c_im_{\rm infl}+h.c.~,  
\label{eq:sdecay}
\end{equation}
and the common, as it turns out, decay width is given by
\begin{equation}
\Gamma=\Gamma_{\theta\rightarrow\bar\psi_{\nu^c_i}
\bar\psi_{\nu^c_i}}=\Gamma_{S\rightarrow\nu^c_i\nu^c_i}=
\frac{1}{8\pi}\left(\frac{M_i}{M}\right)^2m_{\rm infl}~,
\label{eq:gamma}
\end{equation}
provided that the mass $M_i=2\gamma_iM^2/m_P$ of the relevant 
$\nu^c_i$ satisfies the inequality $M_i<m_{\rm infl}/2$. 

\par
To minimize the number of small coupling constants, we assume 
that
\begin{equation}
M_2<\frac{1}{2}m_{\rm infl}\leq M_3=\frac{2M^2}{m_P}~~
({\rm with}~\gamma_3=1),
\label{eq:ineq}
\end{equation}
so that the inflaton decays into the second heaviest right 
handed neutrino superfield $\nu^c_2$ with mass $M_2$. The
second inequality in Eq.(\ref{eq:ineq}) implies that 
$y_Q\leq\sqrt{2N_Q}/\pi\approx 3.34$ for $N_Q\approx 55$.
This gives $x_Q\stackrel{_{<}}{_{\sim}}3.5$. As an 
example, choose $x_Q\approx 1.05$ (bigger values cannot 
lead to an adequate BAU) which yields $y_Q\approx 0.28$. 
From the COBE \cite{cobe} result, we then obtain $M\approx 
4.06\times 10^{15}~{\rm GeV}$, $\kappa\approx 4\times 
10^{-4}$, $m_{\rm infl}\approx 2.3\times 10^{12}~
{\rm GeV}$ and $M_3\approx 1.35\times 10^{13}~{\rm GeV}$.

\par
The `reheat' temperature $T_r$, for the MSSM spectrum, is given 
by \cite{lss}
\begin{equation}
T_r\approx\frac{1}{7}(\Gamma M_P)^{\frac{1}{2}},
\label{eq:reheat}
\end{equation}
and must satisfy the gravitino constraint \cite{khlopov}, 
$T_r\stackrel{_{<}}{_{\sim}}10^9~{\rm GeV}$, for 
gravity-mediated SUSY breaking with universal boundary 
conditions. To maximize the `naturalness' of the model, we 
take the maximal value of $M_2$ (and, thus, $\gamma_2$) 
allowed by this constraint. This is 
$M_2\approx 2.7\times 10^{10}~{\rm GeV}$ ($\gamma_2
\approx 2\times 10^{-3}$). Note that, with this $M_2$, 
the first inequality in Eq.(\ref{eq:ineq}) is well 
satisfied.

\subsection{Baryogenesis via Leptogenesis}
\label{subsec:leptogenesis}

In hybrid inflationary models, it is \cite{lepto} generally 
not so convenient to generate the observed BAU in the usual 
way, i.e., through the decay of superheavy color 
(anti)triplets. Some of the reasons are:

\begin{list}
\setlength{\rightmargin=0cm}{\leftmargin=0cm}

\item[{\bf i.}] 
Baryon number is practically conserved in most models of 
this type. In some cases \cite{bparity}, this 
is a consequence of a discrete `baryon parity' symmetry. In 
the left-right model under consideration, $B$ is exactly 
conserved due to the presence of the R-symmetry.
\vspace{.25cm}
\item[{\bf ii.}] 
The gravitino constraint would require that the mass of 
the relevant color (anti)triplets does not exceed 
$10^{10}~{\rm GeV}$. This suggests strong deviations 
from the MSSM gauge coupling unification and possibly 
leads into problems with proton stability.
\end{list}

\par
It is generally preferable to produce first a primordial 
lepton asymmetry \cite{leptogenesis} which is then partly 
converted into baryon asymmetry by the non-perturbative 
electroweak sphaleron effects \cite{sphaleron}. Actually, 
in the left-right model under consideration as well as in 
many other models, this is the only way to generate the 
observed BAU since the inflaton decays into right handed 
neutrino superfields. The subsequent decay of these 
superfields into lepton (antilepton) $L$ ($\bar L$) and 
electroweak Higgs superfields can only 
produce a lepton asymmetry. It is important to ensure that 
this lepton asymmetry is not erased \cite{turner} by lepton 
number violating $2 \rightarrow 2$ scattering processes 
such as $LL\rightarrow h^{(1)}\,^{*}h^{(1)}\,^{*}$ or 
$L h^{(1)}\rightarrow \bar{L}h^{(1)}\,^{*}$ at all 
temperatures between $T_{r}$ and $100~{\rm GeV}$. This is 
automatically satisfied since the lepton asymmetry is 
protected \cite{ibanez} by SUSY at temperatures between 
$T_r$ and $T \sim 10^{7}~{\rm GeV}$ and, for 
$T\stackrel{_{<}}{_{\sim }}10^{7}~{\rm GeV}$, these 
scattering processes are well out of equilibrium provided 
\cite{ibanez} 
$m_{\nu_{\tau}}\stackrel{_<}{_\sim} 10~{\rm{eV}}$, 
which readily holds in our case (see below). For MSSM 
spectrum, the observed BAU $n_B/s$ is related \cite{ibanez} 
to the primordial lepton asymmetry $n_L/s$ by 
$n_B/s=(-28/79)n_L/s$.

\par
As already mentioned, the lepton asymmetry is produced 
through the decay of the superfield $\nu^{c}_{2}$, which 
emerges as decay product of the inflaton. This superfield 
decays into electroweak Higgs and (anti)lepton superfields. 
The relevant one-loop diagrams are both of the vertex and 
self-energy type \cite{covi} with an exchange of 
$\nu^{c}_{3}$. The resulting lepton asymmetry is 
\cite{neu}
\begin{equation}
\frac{n_{L}}{s}\approx 1.33~\frac{9T_{r}}
{16\pi m_{\rm infl}}~\frac{M_2}{M_3}
~\frac{{\rm c}^{2}{\rm s}^{2}\sin 2\delta
(m_{3}^{D}\,^{2}-m_{2}^{D}\,^{2})^{2}}
{|\langle h^{(1)}\rangle|^{2}~(m_{3}^{D}\,^{2} 
{\rm \ s}^{2}+m_{2}^{D}\,^{2}{\rm \ c^{2}})}~,
\label{eq:leptonasym}
\end{equation}
where $|\langle h^{(1)}\rangle|\approx 174~\rm{GeV}$, 
$m_{2,3}^{D}$ ($m_{2}^{D}\leq m_{3}^{D}$) are the 
`Dirac' neutrino masses (in a basis where they are diagonal 
and positive), and ${\rm c}=\cos\theta$,  
${\rm s}=\sin\theta$, with $\theta$ and $\delta$ being 
the rotation angle and phase which diagonalize the Majorana 
mass matrix of the right handed neutrinos. Note that 
Eq.(\ref{eq:leptonasym}) holds \cite{pilaftsis} provided 
that $M_{2}\ll M_{3}$ and the decay width of 
$\nu^{c}_{3}$ is $\ll(M_{3}^{2}-M_{2}^{2})/M_{2}$, 
and both conditions are well satisfied in our model. Here, 
we concentrated on the 
two heaviest families ($i=2,3$) and ignored the first one. 
We were able to do this since the analysis \cite{giunti} of 
the CHOOZ experiment \cite{chooz} shows that the solar and 
atmospheric neutrino oscillations decouple. 

\par
The light neutrino mass matrix is given by the seesaw 
formula:
\begin{equation}
m_{\nu}\approx-\tilde m^{D}\frac{1}{M}m^{D},
\label{eq:neumass}
\end{equation}
where $m^{D}$ is the `Dirac' neutrino mass matrix and 
$M$ the Majorana mass matrix of right handed neutrinos.
The determinant and the trace invariance of the light 
neutrino mass matrix imply \cite{neu} two constraints 
on the (asymptotic) parameters which take the form: 
\begin{equation}
m_{2}m_{3}=\frac{\left(m_{2}^{D}m_{3}^{D}\right)^{2}}
{M_{2}M_{3}}~,
\label{eq:determinant}
\end{equation}
\begin{eqnarray*}
m_{2}\,^{2}+m_{3}\,^{2}=\frac{\left(m_{2}^{D}\,^{2}
{\rm \ c}^{2}+m_{3}^{D}\,^{2}{\rm \ s}^{2}\right)^{2}}
{M_{2}\,^{2}}+
\end{eqnarray*}
\begin{equation}
\frac{\left(m_{3}^{D}\,^{2}{\rm \ c}^{2}+
m_{2}^{D}\,^{2}{\rm \ s}^{2}\right)^{2}}{M_{3}\,^{2}}+
\frac{2(m_{3}^{D}\,^{2}-m_{2}^{D}\,^{2})^{2}
{\rm c}^{2}{\rm s}^{2}{\cos 2\delta }}{M_{2}M_{3}}~,
\label{eq:trace} 
\end{equation}
where $m_{2}=m_{\nu_\mu}$ and $m_{3}=m_{\nu_\tau}$ are the
(positive) eigenvalues of $m_{\nu}$.

\par
The $\mu-\tau$ mixing angle $\theta_{23}=\theta _{\mu\tau}$ 
lies \cite{neu} in the range
\begin{equation}
|\,\varphi -\theta ^{D}|\leq \theta _{\mu\tau}\leq
\varphi +\theta ^{D},\ {\rm {for}\ \varphi +
\theta }^{D}\leq \ \pi /2~,
\label{eq:mixing}
\end{equation}
where $\varphi$ is the rotation angle which diagonalizes 
the light neutrino mass matrix in the basis where the 
`Dirac' mass matrix is diagonal and $\theta ^{D}$ is the 
`Dirac' mixing angle (i.e., the `unphysical' mixing angle 
with zero Majorana masses for the right handed neutrinos).

\par
We take $m_{\nu_{\mu}}\approx 2.6\times 10^{-3}~
\rm{eV}$ which is the central value from the small angle 
MSW resolution of the solar neutrino problem \cite{smirnov}. 
The $\tau$-neutrino mass is taken to be $m_{\nu _{\tau }}
\approx 7\times 10^{-2}~\rm{eV}$ which is the central value 
implied by SuperKamiokande \cite{superk}. We choose 
$\delta\approx-\pi/4$ to maximize $-n_L/s$. Finally, we 
assume that $\theta^D$ is negligible, so that maximal 
$\nu_\mu-\nu_\tau$ mixing, which is favored by 
SuperKamiokande \cite{superk}, corresponds to 
$\varphi\approx\pi/4$. 

\par
From the determinant and trace constraints and the 
diagonalization of $m_\nu$, we determine the value of 
$m_{3}^{D}$ corresponding to maximal $\nu_\mu-\nu_\tau$ 
mixing ($\varphi\approx\pi/4$) for any given $\kappa$. 
We find that a solution for $m_{3}^{D}$ exists provided 
that $M_2\stackrel{_{<}}{_{\sim }}0.037M_3$. For the 
numerical example in Sec.\ref{subsec:reheat}, we find
$m_{3}^{D}\approx 8.3~{\rm GeV}$ and $m_{2}^{D}
\approx 0.98~{\rm GeV}$. The lepton asymmetry turns out to 
be $n_L/s\approx -2.23\times 10^{-10}$ and, thus, the 
baryogenesis constraint is satisfied. We see that, with not
too `unnatural' values of $\kappa$ ($\approx 4\times 
10^{-4}$) and the other relevant parameters ($\gamma_2
\approx 2\times 10^{-3}$, $\gamma_3\approx 1$), we can 
not only reproduce the results of COBE \cite{cobe} but also
have a successful `reheating' satisfying the gravitino and 
baryogenesis requirements together with the constraints 
from solar and atmospheric neutrino oscillations.

\section{Extensions of SUSY Hybrid Inflation}
\label{sec:extensions}

In trying to apply (SUSY) hybrid inflation to higher GUT 
gauge groups which predict the existence of monopoles, we 
encounter the following problem. Inflation is terminated 
abruptly as the system reaches the critical point on 
the inflationary trajectory and is followed by the 
`waterfall' regime during which the scalar fields 
$\bar\phi$, $\phi$ develop their vevs starting from zero 
and the spontaneous breaking of the GUT gauge symmetry 
occurs. The fields $\bar\phi$, $\phi$ can end up at any 
point of the vacuum manifold with equal probability and, 
thus, monopoles are copiously produced \cite{smooth} via 
the Kibble mechanism \cite{kibble} leading to a 
cosmological catastrophe (see e.g., Ref.\cite{mono}).

\par
One of the simplest GUT models predicting monopoles is the
Pati-Salam (PS) model \cite{ps} based on the gauge group
$G_{PS}=SU(4)_c\times SU(2)_L\times SU(2)_R$. These 
monopoles carry two units of `Dirac' magnetic charge 
\cite{magg}. We will discuss possible solutions 
\cite{smooth,jean} of the magnetic monopole problem of 
hybrid inflation within the SUSY PS model, although our 
mechanisms can be readily extended to other semi-simple 
gauge groups such as the `trinification' group $SU(3)_c
\times SU(3)_L\times SU(3)_R$, which emerges from 
string theory and predicts \cite{trinification} monopoles 
with triple `Dirac' charge, and possibly to simple gauge 
groups such as $SO(10)$.

\subsection{Shifted Hybrid Inflation}
\label{subsec:shifted}

One idea \cite{jean} for solving the monopole problem is 
to include into the standard superpotential for hybrid 
inflation the leading non-renormalizable term, which, as 
explained in Sec.\ref{subsec:susy}, cannot be excluded by 
any symmetries. If its dimensionless coefficient is of 
order unity, this term can be comparable with the trilinear 
coupling of the standard superpotential (whose coefficient 
is $\sim 10^{-3}$). The coexistence of these terms reveals 
a completely new picture. In particular, there appears a 
non-trivial (classically) flat direction along which 
$G_{PS}$ is spontaneously broken with the appropriate Higgs 
fields ($\bar\phi$, $\phi$) acquiring constant values. 
This `shifted' flat direction can be used as inflationary 
trajectory with the necessary inclination obtained again 
from one-loop radiative corrections \cite{dss}. The 
termination of inflation is again abrupt followed by a 
`waterfall' but no monopoles are formed in this transition 
since $G_{PS}$ is already spontaneously broken during 
inflation.

\par
The spontaneous breaking of the gauge group $G_{PS}$ to $G_S$ 
is achieved via the vevs of a conjugate pair of Higgs superfields
\begin{eqnarray}
\bar{H}^c &=& (4,1,2) \equiv \left(\begin{array}{cccc}
                       \bar{u}^c_H & \bar{u}^c_H & 
                       \bar{u}^c_H & \bar{\nu}_H^c\\
                      \bar{d}^c_H & \bar{d}^c_H & 
                      \bar{d}^c_H & \bar{e}^c_H
                      \end{array}\right),\nonumber\\
H^c &=& (\bar{4},1,2) \equiv \left(\begin{array}{cccc}
                       u^c_H & u^c_H & u^c_H & \nu_H^c\\
                       d^c_H & d^c_H & d^c_H & e^c_H
                      \end{array}\right), 
\label{eq:higgs}
\end{eqnarray}
in the $\bar{\nu}_H^c$, $\nu_H^c$ directions. The relevant 
part of the superpotential, which includes the leading 
non-renormalizable term, is
\begin{equation}
\delta W=\kappa S(-M^2+\bar{H}^c H^c)-
\beta\frac{S(\bar{H}^c H^c)^2}{M_S^2}~, 
\label{eq:susyinfl}
\end{equation}
where $M_S\approx 5\times 10^{17}~{\rm GeV}$ is the string 
scale and $\beta$ is taken positive for simplicity. 
D-flatness implies that $\bar{H}^{c} \,^{*}=e^{i\theta}H^c$. 
We restrict ourselves to the direction with $\theta=0$ 
($\bar{H}^{c} \,^{*}=H^c$) containing the non-trivial 
inflationary path (see below). The scalar potential derived 
from $\delta W$ then takes the form
\begin{equation}
V=\left[\kappa(\vert H^c\vert^2-M^2)-\beta\frac{\vert H^c
\vert^4}{M_S^2}\right]^2+2\kappa^2\vert S\vert^2
\vert H^c\vert^2 
\left[1-\frac{2\beta}{\kappa M_S^2}\vert H^c\vert^2
\right]^2.
\label{eq:inflpot}
\end{equation}
Defining the dimensionless variables $w=\vert S\vert/M$, 
$y=\vert H^c\vert/M$, we obtain
\begin{equation}
\tilde{V}=\frac{V}{\kappa^2M^4}=(y^2-1-\xi y^4)^2+
2w^2y^2(1-2\xi y^2)^2, 
\label{eq:vtilde}
\end{equation}
where $\xi=\beta M^2/\kappa M_S^2$. This potential is a 
simple extension of the standard potential for SUSY hybrid 
inflation (which corresponds to $\xi=0$) and appears in a 
wide class of models incorporating the leading 
non-renormalizable correction to the standard hybrid 
inflationary superpotential. 

\par
For constant $w$ (or $|S|$), $\tilde V$ in 
Eq.(\ref{eq:vtilde}) has extrema at  
\begin{equation}
y_1=0,~y_2=\frac{1}{\sqrt{2\xi}},~y_{3\pm}=\frac{1}
{\sqrt{2\xi}}\sqrt{(1-6\xi w^2)\pm\sqrt{(1-6\xi w^2)^2-
4\xi(1-w^2)}}. 
\label{eq:extrema}
\end{equation}
Note that the first two extrema (at $y_1$, $y_2$) are 
$|S|$-independent and, thus, correspond to classically flat 
directions, the trivial one at $y_1=0$ with 
$\tilde{V}_1=1$, and the non-trivial one at  
$y_2=1/\sqrt{2\xi}={\rm constant}$ with 
$\tilde{V}_2=(1/4\xi-1)^2$, which we will use as our 
inflationary path. The trivial trajectory is a valley of 
minima for $w>1$, while the non-trivial one for 
$w>w_0=(1/8\xi-1/2)^{1/2}$, which is its instability 
(critical) point. We take $\xi<1/4$, so that $w_0>0$ and 
the non-trivial trajectory is destabilized before $w$ 
reaches zero (the destabilization is in the chosen direction 
$\bar{H}^{c} \,^{*}=H^c$). The extrema at $y_{3\pm}$, 
which are $|S|$-dependent and non-flat, do not exist for 
all values of $w$ and $\xi$, since the expressions under 
the square roots in Eq.(\ref{eq:extrema}) are not always 
non-negative. These two extrema, at $w=0$, become the SUSY 
vacua. The relevant SUSY vacuum (see below) corresponds to 
$y_{3-}(w=0)$ and, thus, the common vev $v_0$ of 
$\bar{H}^{c}$, $H^c$ is given by
\begin{equation}
(\frac{v_0}{M})^2=\frac{1}{2\xi}(1-\sqrt{1-4\xi}).
\label{eq:v0}
\end{equation}

\par
We will now discuss the structure of $\tilde{V}$ and the
inflationary history in the most interesting range of $\xi$, 
which is $1/4>\xi>1/6$. For fixed $w>1$,
there exist two local minima at $y_1=0$ and 
$y_2=1/\sqrt{2\xi}$, which corresponds to lower potential
energy density, and a local maximum at $y_{3+}$ lying 
between the minima. As $w$ becomes smaller than unity, the 
extremum at $y_1$ turns into a local maximum, while the 
extremum at $y_{3+}$ disappears. The system can freely fall 
into the non-trivial (desirable) trajectory at $y_2$ even 
if it started at $y_1=0$. As we further decrease $w$ below 
$(2-\sqrt{36\xi-5})^{1/2}/3\sqrt{2\xi}$, a pair of new 
extrema, a local minimum at $y_{3-}$ and a local maximum at 
$y_{3+}$, are created between $y_1$ and $y_2$. As $w$
crosses $(1/8\xi-1/2)^{1/2}$, the local maximum at 
$y_{3+}$ crosses $y_2$ becoming a local minimum. At the 
same time, the local minimum at $y_2$ turns into a local 
maximum and inflation along the `shifted' trajectory is 
terminated with the system falling into the local minimum at 
$y_{3-}$ which, at $w=0$, develops into a SUSY vacuum.

\par
We see that, no matter where the system starts from, it 
always passes from the `shifted' trajectory, where the 
relevant part of inflation takes place, before falling 
into the SUSY vacuum. So, $G_{PS}$ is already broken 
during inflation and no monopoles are produced at the 
`waterfall'.

\par
It should be noted that, after inflation, the system 
could fall into the minimum at $y_{3+}$ instead of the one 
at $y_{3-}$. This, however, does not happen since in the 
last e-folding or so the barrier between the minima at 
$y_{3-}$ and $y_2$ is considerably reduced and the decay of 
the `false vacuum' at $y_2$ to the minimum at $y_{3-}$ is 
completed within a fraction of an e-folding before the 
$y_{3+}$ minimum even comes into existence. This transition 
is further accelerated by the inflationary density 
perturbations.

\par
The mass spectrum on the `shifted' trajectory 
can be evaluated \cite{jean}. We find that the only mass 
splitting in supermultiplets occurs in the $\bar{\nu}_H^c$, 
$\nu_H^c$ sector. Namely, we obtain one Majorana fermion 
with ${\rm mass}^2$ equal to $4\kappa^2\vert S\vert^2$, 
which corresponds to the direction
$(\bar{\nu}_H^c+\nu_H^c)/\sqrt{2}$, and two normalized real 
scalars $\mathrm{Re}(\delta\bar{\nu}^c_H+\delta\nu^c_H)$ 
and $\mathrm{Im}(\delta\bar{\nu}^c_H+\delta\nu^c_H)$ with
$m_{\pm}^2=4\kappa^2\vert S\vert^2\mp 2\kappa^2m^2$. Here,
$m=M(1/4\xi-1)^{1/2}$ and 
$\delta\bar{\nu}^c_H=\bar{\nu}^c_H-v$, 
$\delta\nu^c_H=\nu^c_H-v$ with 
$v=(\kappa M_S^2/2\beta)^{1/2}$ being the common value of
$\bar{H}^c$, $H^c$ on the trajectory.

\par
The radiative corrections on the non-trivial trajectory can 
then be found from the Coleman-Weinberg formula \cite{cw} in 
Eq.(\ref{eq:deltav}) and $(\delta T/T)_Q$ and $\kappa$ 
can be evaluated. We find the same expressions as in 
Eqs.(\ref{eq:qa}) and (\ref{eq:kappa}) with $N=2$ ($N=4$) 
in the formula for $(\delta T/T)_Q$ ($\kappa$) and $M$ 
generally replaced by $m$. The COBE \cite{cobe} result
can be reproduced, for instance, with 
$\kappa\approx 4\times 10^{-3}$, which corresponds to 
$\xi=1/5$, $v_0\approx 1.7\times 10^{16}~{\rm GeV}$ (for 
$\beta=1$). The scales 
$M\approx 1.45\times 10^{16}~{\rm GeV}$,
$m\approx 7.23\times 10^{15}~{\rm GeV}$ and the 
`inflationary scale', which characterizes the inflationary 
`vacuum' energy density, $v_{\rm infl}=\kappa^{1/2}m\approx 
4.57\times 10^{14}~{\rm GeV}$. The spectral index $n=0.954$.

\par
The model can be completed by adding the superpotential terms
\begin{equation}
N^2h^2,~\bar{N}^2N^2,~hFF^c,~\bar{H}^c\bar{H}^cF^cF^c,
~G\bar{H}^c\bar{H}^c,~GH^cH^c.
\label{eq:pscouplings}
\end{equation}
In analogy with the model of Sec.\ref{sec:guts}, the first 
two terms are needed for solving the $\mu$ problem via a PQ 
symmetry, the third term represents the Yukuwa couplings, 
with $F=(4,2,1)$, $F^c=(\bar{4},1,2)$ being the `matter' 
superfields, and the fourth term is required for generating 
masses for the $\nu^c$'s (and, thus, for the $\nu$'s) and 
allowing the inflaton to decay. The last two terms, with 
$G=(6,1,1)$, are novel and were introduced \cite{leo} in 
order to give masses to the superfields $\bar{d}^c_H$, 
$d^c_H$ in $\bar{H}^c$, $H^c$.

\par
The model possesses a PQ and a $U(1)$ R-symmetry with 
charges 
\begin{eqnarray*}
PQ:~\bar{H}^c,~H^c,~S,~F^c,~G~(0),~h,~\bar{N}~(1),
~F,~N~(-1);
\end{eqnarray*}
\begin{equation}
R:~\bar{H}^c,~H^c,~h,~\bar{N}~(0),~S,~G~(1),~F,~F^c,
~N~({1/2}). 
\label{eq:pscharges}
\end{equation}
We further impose a $Z_2^{mp}$ symmetry (`matter 
parity'), under which $F$, $F^c$ change sign. Additional 
superpotential terms allowed by the symmetries of the model 
are $FFH^cH^c\bar{N}^2$, $FFH^cH^chh$, 
$FF\bar{H}^c\bar{H}^c\bar{N}^2$, $FF\bar{H}^c\bar{H}^chh$, 
$F^cF^cH^cH^c$. All superpotential terms can be multiplied 
by arbitrary non-negative powers of the combinations 
$\bar{H}^cH^c$, $(\bar{H}^c)^4$, $(H^c)^4$. The symmetry 
which is left unbroken by the vevs of $\bar{H}^c$, $H^c$, 
$\bar{N}$ and $N$ is $G_{S}\times Z_2^{mp}$.

\par
In contrast to the model in Sec.\ref{sec:guts}, $B$ (and 
$L$) violation is present here. It comes from the last 
three additional superpotential terms mentioned above (and 
the combinations $(\bar{H}^c)^4$, $(H^c)^4$) which give 
couplings like $u^cd^cd^c_H\nu^c_H$, $u^cd^cu^c_He^c_H$, 
$ud\bar{d}^c_H\bar{\nu}^c_H$, 
$ud\bar{u}^c_H\bar{e}^c_H$ with appropriate coefficients. 
The couplings $G\bar{H}^c\bar{H}^c$ and $GH^cH^c$ also 
give rise to $B$ (and $L$) violation. Proton can decay via 
effective dimension five operators generated by one-loop 
graphs with two of the $u^c_H$, $d^c_H$ or one of the 
$u^c_H$, $d^c_H$ and one of the $\nu^c_H$, $e^c_H$ 
in the loop. Its lifetime, however, turns out to be long 
enough to make it practically stable.

\par
The `reheating' proceeds as in Sec.\ref{subsec:reheat} with 
$m_{\rm infl}^2=2\kappa^2 v_0^2(1-2\xi v_0^2/M^2)^2$. 
The inflaton again decays into the superfield $\nu^c_2$ 
with mass $M_2=2\gamma_2v_0^2/M_S$, which is evaluated by 
saturating the gravitino constraint. The observed BAU is 
again generated via a primordial leptogenesis from the 
subsequent decay of $\nu^c_2$. The gravitino constraint 
together with the restrictions on the primordial lepton 
asymmetry ($1.8\times 10^{-10}\stackrel{_{<}}{_{\sim }}
-n_L/s\stackrel{_{<}}{_{\sim }}2.3\times 10^{-10}$) 
from the low deuterium abundance constraint 
\cite{deuterium} on the BAU 
($0.017\stackrel{_{<}}{_{\sim }}\Omega_Bh^2
\stackrel{_{<}}{_{\sim }}0.021$) can be satisfied with 
`natural' values of the relevant parameters and in accord 
with the neutrino oscillation data.

\par
A typical solution, for $\kappa=4\times 10^{-3}$ 
($m_{\rm infl}\approx 4.1\times 10^{13}~{\rm GeV}$), 
is $M_2\approx 5.9\times 10^{10}~{\rm GeV}$, $M_3\approx 
1.1\times 10^{15}~{\rm GeV}$ ($\gamma_3=0.5$), 
$m_{\nu_{\mu}}\approx 7.6 \times 10^{-3}~{\rm eV}$, 
$m_{\nu_{\tau}}=8\times 10^{-2}~{\rm eV}$, $m^D_2 
\approx 1.2~{\rm GeV}$, $m^D_3=120~{\rm GeV}$, 
$\sin^2 2\theta_{\mu\tau}\approx 0.87$, and
$n_L/s\approx -1.8\times 10^{-10}$ ($\theta\approx 
0.016$ for $\delta\approx -\pi/3$).

\par
Note that the mass scale $v_0$ is of order $10^{16}
~{\rm GeV}$ which is consistent with the unification of the 
gauge coupling constants of MSSM. Also, the `Dirac' mass 
parameter $m^D_3$, after including renormalization effects 
with large $\tan\beta$ and MSSM spectrum, becomes 
consistent with `asymptotic' Yukawa coupling unification 
which is implied by $SU(4)_c$ (see Ref.\cite{neu}). 
Finally, the $\mu$-neutrino mass turns out to be consistent 
with the large rather than the small angle MSW resolution 
of the solar neutrino problem.

\subsection{Smooth Hybrid Inflation}
\label{subsec:smooth}

An alternative solution to the monopole problem of hybrid 
inflation has been proposed \cite{smooth} some years ago. 
We will present it here within the SUSY PS model of 
Sec.\ref{subsec:shifted}, although it can be applied to 
other semi-simple (and possibly some simple) gauge groups 
too. The idea is to impose an extra $Z_2$ symmetry under 
which $\bar{H}^cH^c\rightarrow -\bar{H}^cH^c$ (say 
$H^c\rightarrow -H^c$). The whole structure of the model 
remains unaltered except that now only even powers of the 
combination $\bar{H}^cH^c$ are allowed in the 
superpotential terms. 

\par
The inflationary superpotential in Eq.(\ref{eq:susyinfl}) 
becomes
\begin{equation}
\delta W=S\left(-\mu^2+\frac{(\bar{H}^cH^c)^2}
{M_S^2}\right), 
\label{eq:smoothsuper}
\end{equation}
where we absorbed the dimensionless parameters $\kappa$, 
$\beta$ in $\mu$, $M_S$. The resulting scalar potential
$V$ is then given by
\begin{equation}
\tilde{V}=\frac{V}{\mu^4}=(1-\tilde\chi^4)^2+
16\tilde\sigma^2\tilde\chi^6, 
\label{eq:smoothpot}
\end{equation}
where we used the dimensionless fields $\tilde\chi=
\chi/2(\mu M_S)^{1/2}$, $\tilde\sigma=
\sigma/2(\mu M_S)^{1/2}$ with $\chi$, $\sigma$ being 
normalized real scalar fields defined by 
$\bar{\nu}_H^c=\nu_H^c=\chi/2$, $S=\sigma/\sqrt{2}$ 
after rotating $\bar{\nu}_H^c$, $\nu_H^c$, $S$ to the 
real axis.

\par
The emerging picture is completely different. The flat 
direction at $\tilde\chi=0$ is now a local maximum with
respect to $\tilde\chi$ for all values of $\tilde\sigma$,
and two `new' symmetric valleys of minima appear 
\cite{smooth,prep} at
\begin{equation}
\tilde\chi=\pm\sqrt{6}\tilde\sigma\left[\left(1+
\frac{1}{36\tilde\sigma^4}\right)^{\frac{1}{2}}-1
\right]^{\frac{1}{2}}.
\label{eq:smoothvalley}
\end{equation}
They contain the SUSY vacua which lie at $\tilde\chi=
\pm 1$, $\tilde\sigma=0$. It is important to note that 
these valleys are not classically flat. In fact, they 
possess an inclination already at the classical level, which 
can drive the inflaton towards the vacua. Thus, there is no 
need of radiative corrections in this case. The potential 
along these paths is given by \cite{smooth,prep}
\begin{eqnarray}
\tilde{V}&=&48\tilde\sigma^4\left[72\tilde\sigma^4\left(1+
\frac{1}{36\tilde\sigma^4}\right)\left(\left(1+
\frac{1}{36\tilde\sigma^4}\right)^{\frac{1}{2}}-1\right)
-1\right]
\nonumber \\
&=&1-\frac{1}{216\tilde\sigma^4}+\cdots,~~{\rm for}
~\tilde\sigma\gg 1.
\label{eq:smoothV}
\end{eqnarray}
The system follows, from the beginning, a particular 
inflationary trajectory and, thus, ends up at a particular 
point of the vacuum manifold leading to no production of 
disastrous magnetic monopoles.

\par
Inflation does not come to an abrupt end in this case since 
the inflationary path is stable with respect to $\tilde\chi$ 
for all $\tilde\sigma$'s. The value $\tilde\sigma_0$ of 
$\tilde\sigma$ at which inflation is terminated smoothly is 
found from the $\epsilon$ and $\eta$ criteria (see e.g., 
Ref.\cite{cosmology}), and the derivatives \cite{prep} of 
the potential along the inflationary path:
\begin{equation}
\frac{d\tilde{V}}{d\tilde\sigma}=192\tilde\sigma^3
\left[(1+144\tilde\sigma^4)\left(\left(1+
\frac{1}{36\tilde\sigma^4}\right)^{\frac{1}{2}}
-1\right)-2\right],
\label{eq:firstder}
\end{equation}
\begin{eqnarray}
\frac{d^2\tilde{V}}{d\tilde\sigma^2}&=&
\frac{16}{3\tilde\sigma^2}
\Biggl\{(1+504\tilde\sigma^4)
\left[72\tilde\sigma^4\left(\left(1+
\frac{1}{36\tilde\sigma^4}\right)^{\frac{1}{2}}
-1\right)-1\right]
\nonumber \\
& &-(1+252\tilde\sigma^4)\left(\left(1+
\frac{1}{36\tilde\sigma^4}\right)^{-\frac{1}{2}}
-1\right)\Biggl\}.
\label{eq:secondder}
\end{eqnarray}

\par
The quadrupole anisotropy of CMBR, $(\delta T/T)_Q$, and 
the number of e-foldings, $N_Q$, of our present horizon 
during inflation can be found from the standard formulae (see 
e.g., Ref.\cite{cosmology}) and using Eq.(\ref{eq:firstder}). 
One advantage of this scenario is that the common vev of 
$\bar{H}^c$, $H^c$, which is equal to $(\mu M_S)^{1/2}$, 
is not so rigidly constrained and, thus, can be chosen equal 
to the SUSY GUT scale $M_G\approx 2.86\times 10^{16}
~{\rm GeV}$. $(\delta T/T)_Q$ can be approximated 
\cite{smooth} as
\begin{equation}
\left(\frac{\delta T}{T}\right)_Q\approx
\frac{1}{\sqrt{5}}
\left(\frac{6}{\pi}\right)^{\frac{1}{3}}
N^{\frac{5}{6}}_Q 
M_G^{\frac{10}{3}}M_P^{-\frac{4}{3}}M_S^{-2}.
\label{eq:smoothqa}
\end{equation}
From the results of COBE \cite{cobe} and taking $N_Q\approx 
57$, we obtain $M_S\approx 7.89\times 10^{17}~{\rm GeV}$, 
$\mu\approx 1.04\times 10^{15}~{\rm GeV}$, which are quite 
`natural'. The relevant part of inflation takes place between 
$\sigma_Q\approx (9N_Q/2)^{1/6}\sigma_0\approx 2.72
\times 10^{17}~{\rm GeV}$ and  $\sigma_0\approx 
(2M_P/9\sqrt{\pi}M_G)^{1/3}M_G\approx 1.08\times 10^{17}
~{\rm GeV}$. 

\par
The inflaton with mass $m_{{\rm infl}}=2\sqrt{2}
(\mu/M_S)^{1/2}\mu\approx 1.07\times 10^{14}~{\rm GeV}$ 
decays again into $\nu^c_2$'s. One can show \cite{prep} 
that the gravitino and leptogenesis constraints can be 
satisfied with `natural' values of the parameters together 
with the restrictions from solar and atmospheric neutrino 
oscillations and $SU(4)_c$ invariance. However, this model 
requires slightly higher $T_r$'s (up to $10^{10}
~{\rm GeV}$), which are perfectly acceptable 
\cite{kawasaki} provided that the branching ratio of the 
gravitino to photons is somewhat smaller than unity and 
$m_{3/2}$ is relatively large (of the order of a few 
hundred GeV).

\section{Conclusions}
\label{sec:concl}
We have shown that, in a wide class of SUSY GUTs, hybrid 
inflation arises `naturally', in the sense that no `tiny' 
coupling constants are needed, the superpotential is 
restricted only by symmetries, and inflation can be 
protected against radiative and SUGRA corrections. 

\par
This inflationary scenario can be readily incorporated in 
concrete SUSY GUT models which simultaneously meet a number 
of other requirements such as the solution of the strong CP 
and $\mu$ problems (via a PQ symmetry), and the generation 
of (seesaw) masses and mixing for light neutrinos.
Moreover, in these concrete models, hybrid inflation is 
followed by a successful `reheating' leading to adequate 
baryogenesis (via a primordial leptogenesis) and satisfying 
the gravitino constraint on the `reheat' temperature 
together with the available solar and atmospheric neutrino 
oscillation data. We give an example of such a SUSY GUT 
model based on the left-right symmetric gauge group 
$SU(3)_c\times SU(2)_L\times SU(2)_R\times U(1)_{B-L}$.

\par
Natural extensions of the standard SUSY hybrid 
inflationary scenario, which still meet all the above 
requirements, can also solve the problem of the possible  
monopole overproduction at the end of inflation. This is 
vital for the application of hybrid inflation to higher GUT 
gauge groups such as the PS group $SU(4)_c\times SU(2)_L
\times SU(2)_R$ or the `trinification' group $SU(3)_c\times 
SU(3)_L\times SU(3)_R$ predicting the existence of 
monopoles.

\section*{Acknowledgement}
This work was supported by the European Union under TMR 
contract No. ERBFMRX-CT96-0090.

\def\anj#1#2#3{(#1),~{\it Astron. J.}~{\bf ~#2},~#3}
\def\apj#1#2#3{(#1),~{\it Astrophys. Journal}~{\bf #2},~#3}
\def\apjl#1#2#3{(#1),~{\it Astrophys. J. Lett.}~{\bf #2},~#3}
\def\baas#1#2#3{(#1),~{\it Bull. Am. Astron. Soc.}~{\bf #2},~#3}
\def\cmp#1#2#3{(#1),~{\it Commun. Math. Phys.}~{\bf #2},~#3}
\def\grg#1#2#3{(#1),~{\it Gen. Rel. Grav.}~{\bf ~#2},~#3}
\def\jetpl#1#2#3{(#1),~{\it JETP Lett.}~{\bf #2},~#3}
\def\jetpsp#1#2#3{(#1),~{\it JETP (Sov. Phys.)}~{\bf #2},~#3}
\def\jhep#1#2#3{(#1),~{\it JHEP}~{\bf ~#2},~#3}
\def\jpa#1#2#3{(#1),~{\it J. Phys.}~{\bf A~#2},~#3}
\def\mnras#1#2#3{(#1),~{\it Mon. Not. Roy. Astr. Soc.}~{\bf #2},~#3}
\def\n#1#2#3{(#1),~{\it Nature}~{\bf #2},~#3}
\def\npb#1#2#3{(#1),~{\it Nucl. Phys.}~{\bf B~#2},~#3}
\def\pl#1#2#3{(#1),~{\it Phys. Lett.}~{\bf #2~B},~#3}
\def\plb#1#2#3{(#1),~{\it Phys. Lett.}~{\bf B~#2},~#3}
\def\pr#1#2#3{(#1),~{\it Phys. Reports}~{\bf #2},~#3}
\def\prd#1#2#3{(#1),~{\it Phys. Rev.}~{\bf D~#2},~#3}
\def\prl#1#2#3{(#1),~{\it Phys. Rev. Lett.}~{\bf #2},~#3}
\def\prsla#1#2#3{(#1),~{\it Proc. Roy. Soc. London}~{\bf A~#2},~#3}
\def\ptp#1#2#3{(#1),~{\it Prog. Theor. Phys.}~{\bf #2},~#3}
\def\spss#1#2#3{(#1),~{\it Sov. Phys. -Solid State}~{\bf #2},~#3}
\def\ibid#1#2#3{(#1),~{\it ibid.}~{\bf ~#2},~#3}
\def\stmp#1#2#3{(#1),~{\it Springer Trac. Mod. Phys.}
~{\bf ~#2},~#3}

\end{document}